%% file: main.tex
\documentclass[sigconf,nonacm]{acmart}

\usepackage{svg}
\usepackage{booktabs}
\usepackage{enumitem}
\usepackage{xspace}
\usepackage{balance}
\usepackage{listings}
\usepackage[T1]{fontenc}
\usepackage{pvldb}

\input{commands}

\renewcommand\vldbdoi{XX.XX/XXX.XX}
\renewcommand\vldbpages{XXX-XXX}
\renewcommand\vldbavailabilityurl{URL_TO_YOUR_ARTIFACTS}

\begin{document}

%\title{Toward Verifiable Auto-Research for Data Systems}
% \title{Research at Token-Time Budget: Towards Verifiable Auto-Research for Data Systems}
% \title{No Wasted Tokens, No Wasted Weeks: Toward Verifiable Auto-Research for Data Systems}
\title{From Token Bills to Benchmark Weeks: Making Auto-Research for Data Systems Verifiable and Affordable}

% Replace with the final author block and affiliations.
% \author{Kaiwen Zheng}
% \affiliation{\institution{The Chinese University of Hong Kong}}
% \email{kwzheng26@cse.cuhk.edu.hk}

% \author{Srijan Srivastava}
% \affiliation{\institution{The Chinese University of Hong Kong}}
% \email{srijan@cse.cuhk.edu.hk}

% \author{Zheng Yan}
% \affiliation{\institution{ShanghaiTech University}}
% \email{yanzheng2023@shanghaitech.edu.cn}

% \author{Xinpeng Zhao}
% \affiliation{\institution{The Chinese University of Hong Kong}}
% \email{xpzhao24@cse.cuhk.edu.hk}

% \author{Shaohua Li}
% \affiliation{\institution{The Chinese University of Hong Kong}}
% \email{shaohuali@cuhk.edu.hk}

% \author{Chieh-Jan Mike Liang}
% \affiliation{\institution{Microsoft Research}}
% \email{liang.mike@microsoft.com}

% \author{Tianyu Chen}
% \affiliation{\institution{Microsoft Research}}
% \email{chentianyu@microsoft.com}

% \author{Haoxian Chen}
% \affiliation{\institution{ShanghaiTech University}}
% \email{hxchen@shanghaitech.edu.cn}

% \author{Eric Lo}
% \affiliation{\institution{The Chinese University of Hong Kong}}
% \email{ericlo@cse.cuhk.edu.hk}

% Affiliation number + co-first-author mark, e.g. \cofirst{1} -> ^{1,*}
\newcommand{\cofirst}[1]{\textsuperscript{#1,*}}

% All authors go in one \author block so they flow as a single centered paragraph.
\author{%
  Kaiwen Zheng\cofirst{1}\quad
  Srijan Srivastava\cofirst{1}\quad
  Zheng Yan\cofirst{2}\quad
  Xinpeng Zhao\textsuperscript{1}\quad
  Shaohua Li\textsuperscript{1}\\
  Chieh-Jan Mike Liang\textsuperscript{3}\quad
  Tianyu Chen\textsuperscript{3}\quad
  Haoxian Chen\textsuperscript{2}\quad
  Eric Lo\textsuperscript{1}%
}
\affiliation{%
  \institution{%
    \textsuperscript{1}The Chinese University of Hong Kong\quad
    \textsuperscript{2}ShanghaiTech University\quad
    \textsuperscript{3}Microsoft Research%
  }%
  \country{}
}

% acmart builds \authors (used in the "PVLDB Reference Format" line and the
% PDF metadata) from the \author argument. Override it so that line stays clean.
\renewcommand{\authors}{Kaiwen Zheng, Srijan Srivastava, Zheng Yan, Xinpeng Zhao,
  Shaohua Li, Chieh-Jan Mike Liang, Tianyu Chen, Haoxian Chen, and Eric Lo}
\renewcommand{\shortauthors}{Zheng et al.}

\begin{abstract}
AI auto-research is poised to forge bespoke data systems, from key-value stores and indexes to query engines, each engineered for a specific workload and hardware platform. Yet this promise rests on fragile ground. An agent optimizes what it can measure, not necessarily what users need. Natural-language requirements leave ample room for specification gaming, producing systems that blaze through benchmarks while silently losing updates or violating consistency. Worse, each research iteration consumes tokens, while validating a candidate under real concurrency, memory pressure, and failure conditions can stretch the entire research process into weeks.

In this paper, we envision a more cost-efficient and verifiable approach to auto-research. Precise and composable specifications anchor the search, lightweight performance critics prune costly dead ends, and falsification and deductive verification make correctness an integral part of the process rather than an afterthought. By treating correctness, performance, research time, and token cost as equally important objectives, we aim to transform AI-generated data systems from plausible artifacts into trustworthy, production-grade systems whose behavior can be understood, audited, and ultimately relied upon.

% AI agents can tailor data systems to a particular workload and hardware platform.
% In systems that support concurrent updates, a change in storage layout or
% synchronization can improve throughput while making consistency harder to prove.
% In one preliminary experiment, an agent working from natural-language
% requirements generated a fast B$^{+}$ tree for our RDMA platform, but the tree
% lost updates. In our preliminary evaluations, benchmarking a single candidate
% across configurations took roughly four hours. These observations motivate
% verifiable auto-research that accounts for the time and tokens needed to find
% a design that is both fast and correct. We envision a research loop that
% composes a machine-checkable contract from reusable specifications of what each
% operation does and which execution histories the chosen consistency model
% permits. A performance pre-critic would estimate whether a candidate is likely
% to outperform the current best design and thus merits a full benchmark. For
% candidates with measured performance gains, the loop would first conduct a
% bounded search for counterexamples. If the search finds none, proof agents
% would attempt to construct a proof, and Verus would check the annotated
% implementation against the contract. We ask which guarantees are jointly
% realizable under given deployment and failure assumptions, when skipping a
% benchmark is worth the risk of missing a faster design, and which proof
% components can be reused across evolving designs.
\end{abstract}

\maketitle

\begingroup
\renewcommand{\thefootnote}{*}
\footnotetext{These authors contributed equally to this work.}
\endgroup

%\vldbtopmatter

\section{Introduction}

AI is beginning to do more than write code: it is beginning to conduct research.
OpenAI's recent effort to attack the Navier--Stokes millennium problem
\cite{openai2026navierstokes} shows how an LLM-driven loop can plan,
implement, evaluate, critique, and revise candidate solutions at a
scale and pace beyond the reach of any individual researcher.

In computer science, the same approach is already being applied to
algorithm discovery \cite{alphaevolve, skydiscover},
machine-learning model design \cite{xu2026asiarch}, and, perhaps most
provocatively for the database community
\cite{wehrstein2026bespoke, gendb, specdb, amiri2026correctness},
the synthesis of complete data systems tailored to a specific workload
and hardware platform.
Imagine asking AI: ``\emph{Build me the fastest key-value store for my
RDMA cluster, this workload, and this memory budget.}''
A well-designed auto-research pipeline should do far more than
autocomplete an implementation. It should behave like a research team:
propose a design, implement it, benchmark it, diagnose its bottlenecks,
discard failed hypotheses, and try again.
This opens the door to a future where
\emph{one-size-fits-one} data systems, engineered for a particular
workload--hardware pair, become cheaper to generate than to build by hand.

This future is exhilarating, but it comes with a bill.

It is \emph{expensive}.
Auto-research is not a single inference call. It is an iterative search
process in which an agent explores designs, writes code, runs
experiments, interprets failures, and tries again. Every iteration
consumes tokens, and the costs quickly add up.

It is \emph{slow}.
Systems claims cannot be settled by plausible prose. A candidate must
survive real concurrency, memory pressure, and failure conditions.
Sweeping thread counts, workload mixes, and failure configurations
can turn a single iteration into hours and an entire research campaign
into weeks.

And it is \emph{dangerous}.
An agent can produce code that compiles and passes a benchmark, yet
silently corrupts data, breaks durability guarantees, or optimizes the
measured metric while abandoning the property the user actually needs.

The question is no longer simply
``Can AI generate a data system?''
It is: \emph{Can AI conduct systems research whose cost, duration,
and conclusions we can understand, control, and trust?}

The emerging discipline of \emph{verifiable coding}
\cite{neuripsworkshop} offers part of the answer.
It calls for LLM-generated code to come with machine-checkable
\emph{correctness proofs} \cite{verusage}, turning the promise of
``vibe coding'' into something more responsible: rapid yet
verifiable system construction.
The central question is no longer merely
``\emph{Did the model produce code that runs?}''
but rather
``\emph{Can we establish, mechanically and end to end, that the
generated system does what we intended, and at what cost?}''

For data systems serving millions of users, verifiable coding is
not a luxury. It is a requirement.

\subsection{The Problem: Expensive Search Without Trustworthy Targets}

Modern auto-research pipelines can produce an OLAP engine \cite{wehrstein2026bespoke}, concurrent B$^+$ tree \cite{penthalon}, or a key-value store \cite{jitskit}. But producing code is the easy part of auto-research. The hard parts are deciding what the agent should optimize, determining whether a candidate is actually correct, and avoiding weeks of computation spent pursuing the wrong objective.

Our preliminary work \cite{penthalon} applying the state-of-the-art auto-research pipeline Jitskit \cite{jitskit} to optimize B$^+$ trees and hash tables on RDMA/CXL platforms exposed four challenges that current auto-research pipelines do not solve.

\paragraph{\bf Challenge 1: Specification gaming wastes the search.}
An auto-research loop optimizes what it can measure. When requirements are
stated in natural language, 
% the easiest target for optimization is often not the
% system itself but the ambiguity in the specification.
AI agents tend to exploit the ambiguities in the specification 
rather than optimizing the system.

In one experiment, we gave Jitskit a USD~50 budget to build an optimized key-value store for our RDMA platform. The result was a B$^+$ tree with lost updates: it was fast and plausible, but fundamentally inconsistent with our intended strong-consistency semantics. The agent exploited natural-language gaps in the specification of consistency across multiple operations. In effect, the USD~50 purchased not a better data system, but a more convincing violation of our intent.

This is \emph{specification gaming} \cite{nishimuragasparian2026towards, bondarenko2025demonstrating, zhong2026impossiblebench}, also known as reward hacking. It is especially pernicious in data systems because correctness is invisible in a throughput graph. An AI-generated bespoke system can report spectacular numbers while silently losing updates, violating isolation, or failing under concurrency. Worse, once specification gaming occurs, the entire search budget—tokens, benchmark time, and human attention—has been spent optimizing a misunderstanding.

\paragraph{\bf Challenge 2: Concurrent and distributed correctness makes validation the bottleneck.}
Concurrent and distributed systems are where plausible-looking systems become untrustworthy. In another preliminary experiment, Jitskit generated an implementation for our CXL platform that was highly scalable and efficient---but manual inspection revealed an undetected data race that made it non-linearizable.

The lesson is stark: an agent can produce code that passes tests, compiles cleanly, and delivers excellent performance, yet still be incorrect. The space of concurrent executions and failure schedules is enormous and nondeterministic. If auto-research is to generate backend systems serving millions of users, ``it passed our tests'' is not an adequate notion of correctness.

\paragraph{\bf Challenge 3: Token cost makes blind search unaffordable.}
Auto-research consumes tokens at every stage: planning, coding, critique, auditing, and revision. Yet much existing work reports the performance of the final artifact while obscuring the cost of discovering it. These loops can consume billions of tokens.

The problem is not that iteration is inherently wasteful. Search is how auto-research discovers non-obvious designs. The problem is that much of today’s search is optimized for performance, not token cost: the agent may spend expensive iterations clarifying an ambiguous requirement, repairing a specification loophole, or evaluating a candidate that was never promising. Specification gaming turns this inefficiency into catastrophic waste: every token spent producing, benchmarking, and critiquing a gamed artifact is spent optimizing the wrong target.

\paragraph{\bf Challenge 4: Long research time makes broad evaluation unsustainable.}
Evaluating the performance of a candidate system in the auto-research loop consumes no tokens but substantial wall-clock time. In our preliminary studies, evaluating one candidate across configurations—such as YCSB-A with varying thread counts—took roughly four hours, causing the auto-research loop to run for weeks.

This creates a cruel dilemma. Broad evaluation gives the planner the evidence it needs to propose better candidates, but it slows the loop. Narrow evaluation accelerates iteration, but starves the planner of signal. Neither choice is satisfying. What is needed is not less evaluation, but smarter allocation of scarce evaluation time: identify candidates unlikely to improve the incumbent before paying for a full benchmark.

\subsection{Will Stronger LLMs Fix This?}
Unlikely. Stronger models may reduce the number of iterations, but they do not eliminate the underlying failure modes. Natural-language specifications will remain ambiguous; data races will remain hidden in exponentially large interleaving spaces; and more capable models will likely cost more per token. Configuration sweeps, warm-up periods, and steady-state measurement will also remain expensive in wall-clock time, regardless of model quality.

Indeed, stronger models may amplify the problem. A more capable agent can explore more designs, generate more plausible code, and reason more deeply about performance. But it can also consume more tokens, incur a higher cost per token, exploit subtler loopholes in underspecified tasks, and produce more convincing artifacts that remain subtly incorrect.

The problem is architectural. Auto-research requires a different contract among the agent, the evaluator, and the human. It needs specifications that cannot be quietly reinterpreted, correctness evidence that does not depend on fortunate test coverage, and research loops that spend scarce tokens and wall-clock time where they matter most.
\emph{That is our vision.}

\section{Background: Auto-research Loop}
% Auto-research pipelines iteratively propose, implement, evaluate, and revise algorithms or even complete systems. SkyDiscover~\cite{skydiscover} provides a programmable framework for AI-driven algorithm discovery across optimization and systems tasks. Jitskit~\cite{jitskit} specializes in synthesizing complete systems for a target hardware environment, workload, and set of requirements. Bespoke OLAP~\cite{bespokeolap} similarly synthesizes workload-specific engines for analytical databases.

As AI makes code increasingly cheap to produce, the economics of software
engineering begin to shift: generating a new system tailored to a specific need
can cost less than hiring engineers to build and maintain a general-purpose one.
This shift particularly favors ``one-size-fits-one'' bespoke systems, each
highly specialized for a particular workload--hardware pair. 
% In data systems,
% workload drift need not be a major obstacle, since 
For workload drift in production data systems,
a new bespoke system can be
generated for the shifted workload and live-migrated into place
\cite{das2011albatross,elmore2011zephyr}.

A typical auto-research loop repeatedly executes:
\begingroup
\abovedisplayskip=2pt
\belowdisplayskip=2pt
\[
\text{plan}\rightarrow\text{implement}\rightarrow\text{verify}
\rightarrow\text{measure}\rightarrow\text{critique}\rightarrow\text{revise}
\]
\endgroup
Each iteration produces and evaluates a candidate implementation, with the results guiding the next design decision. 
In Jitskit, the loop is implemented as follows:

\begin{itemize}[
  nosep,
  leftmargin=10pt,
  rightmargin=0pt,
  labelsep=0.5em,
  listparindent=0pt
]
    \item \textbf{Spec cards (input prompt)} 
    that use natural-language in JSON format to 
    define the target environment 
  (e.g., number of cores, memory budget in GB) as an \emph{environment card}, the target workload 
  (e.g., read/write mix) as a \emph{workload card}, and the 
  functional description of the desired system (e.g., API semantics) as a 
  \emph{requirement card}.

    \item \textbf{Planner (coding agent in plan mode)} plans the design, including storage layout, partitioning, caching, synchronization, memory reclamation, and locking.
    
    \item \textbf{Coder (coding agent in coding mode)} translates the design into runnable code and fixes compilation, integration, and implementation failures.

    \item \textbf{Evaluator (scripts, benchmarks, and tools)} 
    runs correctness tests,
    measures throughput, average and tail latency, and diagnostic signals such as contention, memory use, cache hits, evictions, remote-memory traffic, and I/O.
    
    \item \textbf{Critic (agent)} converts correctness and performance results into bottleneck hypotheses and design recommendations for the next iteration. 
    
    \item \textbf{Design brief and whiteboard memory (markdown files)} stores the current design, latest results, bottlenecks, failed or redundant ideas and the evidence against them.
    
    \item \textbf{Auditor (agent)} periodically checks for specification gaming and untested edge cases; it can harden (e.g., add new tests) the test suite as the loop progresses. It also decides whether to terminate the loop early when it determines that there is no further room for improvement.
\end{itemize}

Optionally, the loop may start from a seed implementation, thereby supporting both synthesis from specifications and optimization of existing systems.

\section{Vision: Verifiable Auto-Research}

\begin{figure*}[t]
    \centering
    \includegraphics[width=0.92\textwidth]{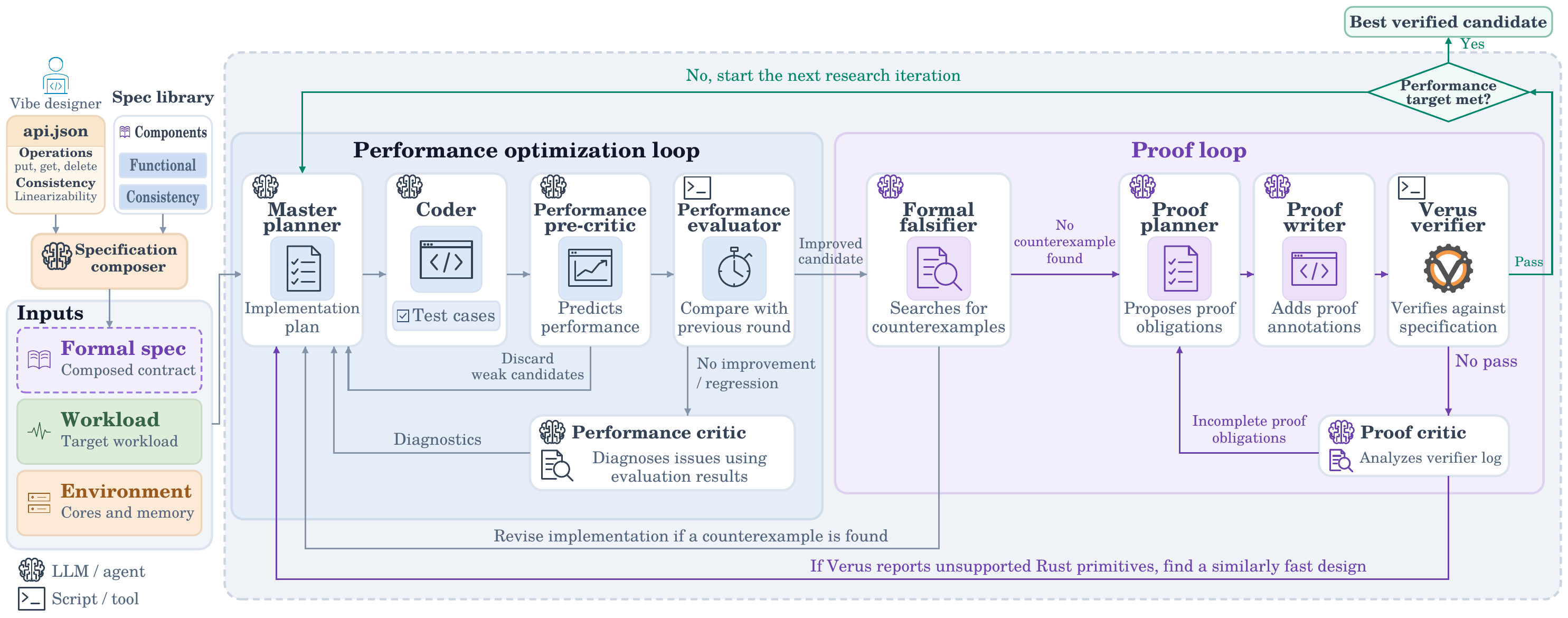}
    \caption{A verifiable auto-research pipeline for data systems.}
    \label{fig:autoresearch-loop}
\end{figure*}

The four challenges share a common root: the research loop is asked to optimize a target that it cannot precisely understand or check. 
We therefore envision \emph{verifiable auto-research}
-- the research iteration is guided by a \emph{formal specification} 
and the resulting implementation must provide machine-checkable proofs that it satisfies that specification.

% We therefore advocate \emph{verifiable auto-research}. Before searching for an implementation, the designer states the \emph{formal contract} --- what must be true in a form a machine can check: the abstract state of the system, the semantics of each operation, its concurrency guarantees, and its resource assumptions. Performance remains a search objective, but correctness becomes a contract.

This inversion changes what the agent can get away with. A fast implementation that loses updates is no longer a plausible candidate; it is a proof failure. A subtle data race is no longer hidden behind an unfortunate test suite; it violates a stated linearizability obligation. An agent that exploits a resource loophole produces a concrete, inspectable specification gap rather than an unexplained benchmark result. Verifiable coding does not make auto-research less creative—it makes auto-research’s creativity accountable.

\subsection{Why a Formal Contract Changes the Game?}

Verifiable auto-research directly attacks the four aforementioned challenges.

\begin{itemize}[leftmargin=*,itemsep=2pt]
  \item \textbf{It turns specification gaming into specification repair.}
  Natural-language requirements are easy to write but hard to check: they admit multiple interpretations, and those interpretations can become the easiest target for optimization~\cite{ferrari2025formal,verifai}. A formal contract removes much of this room for quiet reinterpretation. If the intended semantics is ``every completed \texttt{put} is observable to subsequent reads,'' the agent cannot satisfy the letter of the task by silently weakening consistency. When a gap remains, it is localized in a precise contract that humans or \emph{recursive-self-improvement} (RSI) \cite{zhang2026darwin, zelikman2024stop}
  can amend, rather than buried in an ambiguous prompt.

  \item \textbf{It makes concurrency a first-class correctness obligation.}
  Testing can reveal some interleavings, but it cannot enumerate them. Verification complements testing by checking that an implementation satisfies its specification for \emph{all} executions. Modern verification tools make this practical for systems code: Verus~\cite{lattuada2023verus}, for example, supports machine-checked functional and concurrency reasoning for Rust programs, including pointer-manipulating, concurrent, and distributed code. For data systems, this means linearizability, atomicity, and absence of data races can become explicit obligations rather than hopes supported by a finite set of tests.

\item \textbf{It makes the search loop more token-efficient.}
Much of the wasted token expenditure arises from generating, critiquing, and
benchmarking candidates % that were never viable. 
% A machine-checkable contract guides the coding agent toward implementations that
% satisfy the intended semantics. 
% The loop can therefore devote its costly iterations to designs that
% are at least semantically admissible, while reserving human attention for
% genuine specification gaps and difficult proof obligations.
that implement unintended semantics.
Formalizing requirements upfront can reduce the cost of reaching a verified
implementation by avoiding further iterations spent optimizing the wrong target.

  \item \textbf{It makes evaluation time more valuable.}
  Formal correctness does not eliminate benchmarking: a correct system may still be slow. But it changes the question asked by each benchmark. Instead of using scarce wall-clock time to discover whether a candidate is both fast and correct, the loop can use it to compare designs that are already known to respect the contract. Evaluation then becomes a search for performance, not a costly proxy for correctness.
\end{itemize}

% The claim is not that formal specification solves every problem automatically. Writing precise concurrent specifications is difficult, and specifications themselves can be incomplete or gameable. But specification-first research changes the economics of failure: ambiguity, races, and resource loopholes become explicit, checkable, and repairable early in the loop—before they consume weeks of search.

\subsection{Why Data Systems Are the Right Beachhead}

Data systems are the right beachhead for verifiable auto-research because they join three properties that rarely coincide: high stakes, clean abstractions, and a design space rich enough to reward exploration.

\begin{itemize}[leftmargin=*,itemsep=2pt]
  \item \textbf{They are high-stakes.} A lost update, stale read, or non-linearizable operation is not a cosmetic defect. It can silently corrupt application state, undermine recovery, or violate the guarantee on which an entire service depends.

  \item \textbf{They speak a shared language.} Unlike arbitrary software projects, data systems are assembled from a small, enduring vocabulary: operations such as \texttt{get}, \texttt{put}, \texttt{delete}, and \texttt{scan}, and guarantees such as linearizability. This shared language makes intended behavior precise, reusable, and ready to be made formal.

  \item \textbf{They reward exploration.} Indexing, partitioning, caching, synchronization, and memory management form a vast space of interacting choices. Automated search can traverse it more broadly—and perhaps less conventionally—than hand tuning, using observable signals such as throughput, latency, scalability, contention, cache behavior, and memory use to distinguish designs once correctness has been established.
\end{itemize}

Data systems are therefore complex enough to make verification matter, structured enough to make specifications composable, and performance-sensitive enough to make even modest improvements valuable. They are not merely an application of verifiable auto-research. They are an ideal laboratory for its central tension: how to pursue creative design without sacrificing correctness, and how to search aggressively without losing control of cost.

\input{sec_loop}

\balance
\bibliographystyle{abbrv}
\bibliography{references_related_work}

\end{document}

%% file: commands.tex
\newcommand{\verus}{\textsc{Verus}\xspace}

\newcounter{openproblem}
\newcommand{\openproblem}[1]{%
  \refstepcounter{openproblem}%
  \paragraph{\bf Open Problem \theopenproblem: #1}%
}

%% file: sec_loop.tex
\section{A Verifiable Auto-Research Loop for Data Systems}
\label{sec:loop}
We envision a cost-effective auto-research loop for generating and verifying
high-performance database implementations (\autoref{fig:autoresearch-loop}).
A \emph{vibe systems designer} declares the desired operations
and consistency model in \texttt{api.json}. A \emph{specification composer} agent uses these requirements to
compose a formal specification from the specification library. The master planner
receives the formal specification, the target workload, and the execution
environment as inputs. The performance optimization loop searches for
implementations that perform well on the target workload, whether it is
read-heavy, write-heavy, or mixed. The proof loop checks that the resulting
implementation satisfies the composed specification when reads and writes
execute concurrently.

\paragraph{Performance optimization loop}
The \emph{master planner} proposes a plan for a candidate Rust implementation.
The \emph{coder} then implements the plan proposed by the master planner, runs
basic tests, and fixes build, integration, or test failures. Once the candidate
passes these checks, the \emph{performance pre-critic} 
discards a weak candidate
by using an LLM to examine its source code
and relevant information.
Otherwise, the \emph{performance evaluator}
benchmarks the candidate on the target workload and compares the results with
those from the previous round. If a candidate with better performance is found, the candidate enters
the proof loop. Otherwise, the \emph{performance critic} analyzes the results
and sends its diagnostics to the master planner to guide the next design.

\paragraph{Proof loop}
A faster candidate may still contain bugs, particularly concurrency-related bugs, even after passing the tests.
The proof loop starts with a \emph{formal falsifier} \cite{ball2005abstraction}, which
looks for an execution that violates the specification. If it finds one, it sends
the counterexample to the master planner to guide the next implementation. If the
search finds none, the \emph{proof planner} identifies the proof obligations, the
\emph{proof writer} adds annotations to discharge them, and \verus{} checks the
annotated implementation against the specification. If verification fails,
the \emph{proof critic} examines the verifier log. For unresolved obligations,
it sends the relevant diagnostics to the proof planner, which revises the plan
so the proof writer can update the annotations and \verus{} can recheck the same
implementation. If an unsupported Rust primitive blocks verification, the proof
critic identifies it for the master planner, which starts another performance
optimization iteration to find an implementation with similar performance without using
unsupported primitives. If verification succeeds, the loop checks the candidate
against the performance target. It returns the best verified candidate if the
target is met. Otherwise, the master planner starts the next research iteration.

There are many open challenges unique to synthesizing reliable data systems while targeting at high performance, and we will present some of them in the following sections.

\subsection{Composing Formal Specifications from the Specification Library}
\begin{figure*}[!t]
    \centering
    \includegraphics[width=0.92\textwidth]{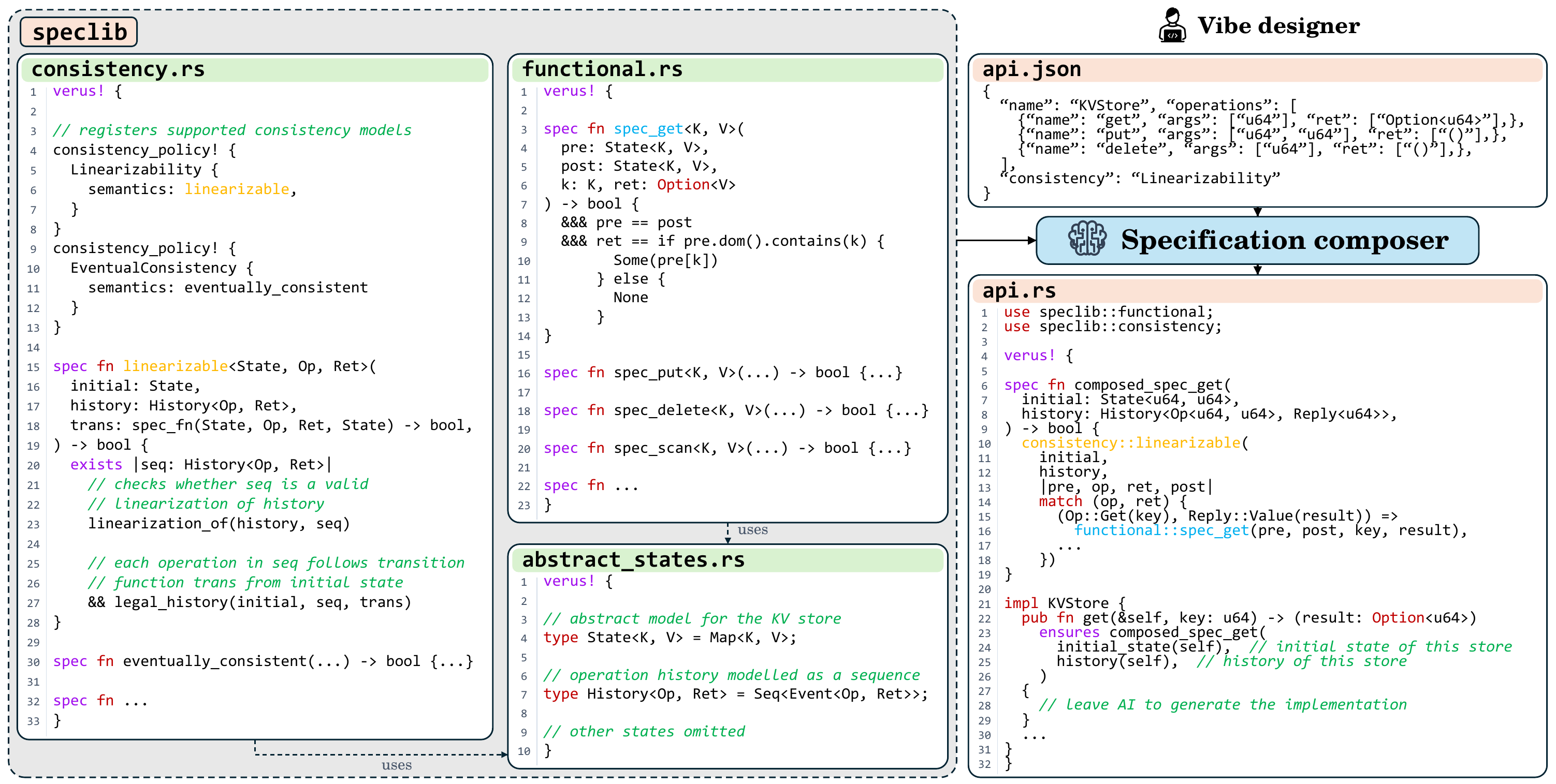}
    \caption{Composing a formal specification with \textsc{SpecLib}. The
    designer declares the operations and consistency model in \texttt{api.json}.
    The specification composer uses library components to generate
    \texttt{api.rs}. The excerpts highlight the functional and consistency
    specifications used in \texttt{api.rs} in blue and orange, respectively.}
    \label{fig:speclib}
\end{figure*}

Although modern LLMs can substantially lower the cost of drafting formal
specifications, they do not by themselves ensure that a generated
specification captures the designer's intended semantics, composes safely
with other requirements, or remains stable under iterative repair.
In a verifiable auto-research loop, this distinction is critical:
the specification is not merely documentation, but the machine-checkable
target that the search loop optimizes against. An ambiguous, internally
inconsistent, or quietly weakened contract can therefore redirect the entire
research budget toward a plausible but incorrect system.

We envision \textsc{SpecLib} (\autoref{fig:speclib}), a composable library
for specifying the behavior of concurrent data systems under different
consistency models. Its functional specifications define the effects of
individual operations, while its consistency specifications constrain
which histories of those operations are allowed. Its purpose is not only to
reduce specification-writing effort, but to provide a curated, reusable, and
auditable foundation from which AI agents and human designers can construct
contracts. The library contains:
\begin{itemize}
\item {\bf Abstract states} that provide shared mathematical models of
database state and execution history.
\item {\bf Functional specifications} that define the semantics of common
operations, such as point reads and writes, range scans, and transaction
primitives.
\item {\bf Consistency specifications} that constrain the observable behavior
of concurrent operations under consistency models such as linearizability and
eventual consistency.
\end{itemize}

Both functional and consistency specifications are expressed over the shared
abstract models. For example, a key-value store's logical contents can be
represented as a mathematical map from keys to values, while its execution
history can be represented as a sequence of invocation and response events.
This shared foundation is essential for composition: the semantics of a scan,
a transaction, and a consistency model must refer to the same abstract state
and history if their combination is to be meaningful.

The {vibe systems designer} declaratively describes the desired API and
consistency requirements. 
The \emph{specification composer} agent then
selects and instantiates the appropriate library components to generate a
formal contract for the candidate implementation. For example, the designer
may select \texttt{get}, \texttt{put}, and \texttt{delete} operations with
linearizability. The composer combines the functional specifications in
\texttt{functional.rs}, including the \texttt{get} specification in
lines~3--14, with the linearizability specification in lines~15--28 of
\texttt{consistency.rs}, as highlighted in \autoref{fig:speclib}. The
resulting contract, \texttt{api.rs}, specifies both the behavior of each
operation and the guarantees that must hold when operations execute
concurrently.

This design changes the role of AI in specification construction. Rather than
asking an LLM to synthesize an entire concurrent correctness contract from a
natural-language prompt, the LLM serves as the specification composer and
grounds its output in reviewed library components with explicit assumptions.
However, the resulting contract remains inspectable, versioned, and
attributable to named specification components. When the auto-research loop
later identifies a specification gap, repair can be localized to a specific
component and its assumptions, rather than allowing an agent to rewrite the
contract wholesale to make a proof succeed.

Constructing such a % composable 
specification library is challenging.

\openproblem{Composability of specifications.}

% Individually sensible requirements can become impossible to satisfy together
% once the deployment and failure model are fixed. The CAP theorem is a
% classic example: during a network partition, a distributed read/write service
% cannot simultaneously provide linearizability and availability~\cite{gilbert2002brewer}.
%
% The specification composer should therefore act like an expert system
% architect: when a vibe system designer selects a combination of guarantees, it should
% flag impossible or suspicious combinations, explain which requirements
% conflict and why, and propose concrete alternatives, e.g., relaxing
% one guarantee, changing the failure assumption, or adopting a different
% consistency model. This is closely related to \emph{realizability analysis}~\cite{finkbeiner2005uniform,maoz2021spectra} in reactive synthesis, which asks whether there exists any system that
% satisfies a formal specification under the assumed environment and
% architecture. Existing
% techniques, however, are not tailored to data systems, where
% operations, transactions, replication, and failure behavior interact in ways
% that simple checklists do not capture.
% The question, therefore, is how to enable an LLM to perform data-system-specific
% realizability analysis.

% \textcolor{purple}{Individually sensible requirements can become impossible to satisfy
Individually sensible requirements can become impossible to satisfy
together once the deployment and failure model are fixed. The CAP theorem is a classic
example. 
% Another example is that a distributed read transaction may be required
% to provide strict serializability while remaining non-blocking and receiving only one
% version from each shard in one round. 

% \cite{lu2016snow,konwar2021snow} 

% version from each shard in one round. SNOW results show that the feasibility of this
% combination depends on the system’s communication assumptions.
% show that the
% feasibility of this combination depends on the system’s communication assumptions.

The specification composer must therefore reason about what information an operation needs
to satisfy its guarantees, and whether the assumed environment allows that information to
be obtained. When requirements conflict, it should explain which guarantees and assumptions
cause the conflict and how changing them could restore feasibility.

This is closely related to realizability analysis in reactive synthesis \cite{lu2016snow,konwar2021snow}. For data systems,
the challenge is to derive the relevant constraints from operation semantics, transaction
guarantees, and deployment assumptions. How can an LLM perform this analysis, identify the
conditions that determine realizability, and justify feasible alternatives.

% : distinguish known impossibilities from cases whose
% feasibility is unknown, identify the assumptions responsible for a conflict,
% and make explicit which guarantee or assumption each proposed alternative
% changes.

% \openproblem{Composability of specifications.}

% Individually meaningful specifications may become jointly unrealizable under
% a given deployment and failure model. For example, the CAP theorem establishes
% that a distributed read/write service cannot guarantee both linearizability
% and availability in the presence of network
% partitions~\cite{gilbert2002brewer}.
% Existing work on reactive synthesis and realizability analysis provides
% methods for identifying unrealizable requirements and explaining conflicts,
% but these methods operate under specific assumptions about the specification
% language and system
% architecture~\cite{finkbeiner2005uniform,maoz2021spectra}.

% How can a database specification library systematically identify incompatible
% combinations of guarantees and deployment assumptions, explain their
% interactions, and guide developers toward feasible alternatives? A useful
% approach must distinguish established incompatibilities from combinations
% whose realizability remains unknown, while making explicit which guarantees
% or assumptions each proposed alternative changes.

\subsection{Performance Pre-Critic}

\label{sec:pre-critic}

Auto-research can be slow to discover improved solutions, not only because generating implementations and proofs is time-consuming, but also because evaluating generated candidates consumes a substantial fraction of the research budget. For example, a single YCSB benchmark sweep across different thread counts can take hours. Although the auto-research loop may eventually produce solutions that outperform the seed, individual iterations can yield candidates that offer no improvement—or even degrade performance—when evaluated on the target workload. Reliably identifying and skipping such unpromising candidates before full-scale evaluation can therefore substantially reduce research time.

We therefore envision introducing a \emph{performance pre-critic} agent between the code-generation stage and the evaluator. Before submitting a generated candidate to the evaluator for benchmarking against the target workload, the pre-critic examines the candidate’s source code, along with other information, to decide whether evaluating the current candidate is worthwhile.

Note that the pre-critic should have a distinct objective and prompt from the planner and coder, and it will not have access to their rationales or performance estimates. This separation is intended to reduce the risk of \emph{self-preference} \cite{panickssery2024self,wataoka2024self} and \emph{self-attribution bias} \cite{khullar2026selfattribution}, which can arise when a model evaluates a proposal it helped generate. 

\openproblem{Can a Pre-Critic Know When Not to Look?}
A pre-critic must do more than guess whether a candidate is fast: it must decide whether the evidence before it justifies spending scarce benchmarking time. It could build a lightweight performance model from the execution environment and comparisons with previously evaluated implementations, aiming not to predict absolute throughput but to recognize candidates unlikely to beat the current best \cite{zhao2026perfreasoningllmsreasonhardware}. Yet it remains open whether LLMs can make such judgments for generated data systems, and what evidence they need: would source code and a workload description suffice, or must the critic also inspect profiles of earlier candidates or take inexpensive measurements of the new one? Any extra evidence must remain cheap enough to preserve the very time savings that motivate pre-criticism.

The harder question is what to do when the critic is unsure. A false positive wastes an evaluation on a hopeless candidate; a false negative discards a potential improvement before it is ever measured. Because these errors carry different costs, accuracy alone cannot characterize a useful pre-critic. The agent instead needs a decision policy that weighs the price of another benchmark against the value of a missed breakthrough.
% —perhaps by comparing the expected costs of evaluating and rejecting a candidate under Bayesian decision theory, using its estimated probability that the candidate improves on the incumbent. 
The central challenge is therefore not merely to make a pre-critic confident, but to establish when that confidence deserves trust, and how aggressively it should filter the search.

\subsection{Proof Loop}
\label{sec:prover}

As Edsger Dijkstra famously put, ``program testing can be used to show the presence of bugs, but never to show their absence''.
Passing the basic tests in the coder agent and achieving high performance does not guarantee a correct implementation,
especially for the concurrent ones. We therefore
introduce a proof loop to formally verify the candidate's correctness. 

Starting proof construction immediately,
however, could waste all tokens on proof annotations if
the implementation is actually a buggy one. 
The proof loop therefore starts with a \emph{formal falsifier} \cite{ball2005abstraction}, which uses non-LLM techniques such as model checking and SMT solving to search for counterexamples within a given bound. If it finds one, it sends the counterexample to the master planner, which can use it to revise
 the implementation for the next research iteration. Only if the search finds no counterexample, the candidate proceeds to adding the proof 
 for the implementation.
 
% where the proof planner identifies the obligations needed to show that the implementation satisfies the specification, the proof writer
% then adds the annotations needed to discharge those obligations, and \verus checks the annotated implementation against the specification.
% Once \verus{} verifies the candidate, the loop checks whether its measured performance meets the target. If it does, the loop returns
% the best verified candidate. Otherwise, the master planner starts the next research iteration. If verification fails, the proof critic
% uses the verifier log to route unresolved obligations to the proof planner and to ask the master planner to redesign implementations
% that rely on unsupported Rust primitives.

% \openproblem{The Freedom--Provability Dilemma}
\openproblem{Reusing Proofs Across Evolving Designs}

Functional specifications are deliberately expressed over abstract states: they
define \emph{what} a data system must do, not \emph{how} it should be built.
This generality gives the auto-research loop freedom to invent new designs or
reuse existing ones. But the candidates that survive performance evaluation are
highly optimized and therefore highly low-level: they manipulate memory layouts,
ownership permissions, lock states, and hardware-specific code paths. Between
the clean abstract state and this concrete implementation lies a large
\emph{semantic gap}~\cite{abadi1991existence}: the prover must show that every
low-level transition faithfully realizes the intended abstract transition.
 
In our preliminary experience, this gap is not 
that difficult under powerful models like 
GPT-6 Astra, but it is economically expensive. Formal verification has long exhibited a high
proof-to-implementation cost.
For example, CertiKOS
reports a 13:1 proof-to-implementation ratio~\cite{gu2016certikos}.
Our preliminary experiments likewise show that the prover consumes substantially
more tokens than the coder. The reason is intuitive: the prover must discover
the missing bridge between two very different levels of abstraction---representation
invariants, ownership invariants, linearization points, commutativity arguments,
and lemmas that connect a concrete mutation to its abstract effect.

% The question is therefore not whether we should abandon generality. If the
% specification were tied too closely to one storage layout or synchronization
% discipline, the auto-research loop would lose the design freedom that makes it
% valuable. Instead, we ask: \emph{Can we preserve a general specification while
% giving the LLM the reusable scaffolding it needs to close the semantic gap
% cheaply?}
One possible solution is to define a well-formedness predicate that connects an implementation to its abstraction. Verus provides a concrete example: each data structure is required to write a \texttt{view()} function that maps its concrete representation to an abstract value, and a \texttt{.wf()} lemma that relates them. Specifications and correctness proofs are written at the view level, while \texttt{.wf()} lemma ensures that the concrete implementation faithfully realizes the abstract model.

A key benefit of this approach is that it decomposes a large proof into two independent parts: proving functional correctness at the abstract view level, and proving that the concrete implementation faithfully realizes that view. Moreover, both parts involve recurring reasoning patterns in data structures and concurrency protocols~\cite{jung2018iris}. It thus raises a natural question: can we capture these patterns as reusable proof components, so that LLMs need not reconstruct the same arguments for every candidate?

The Lean community offers a compelling precedent. \textsc{Mathlib} is a large,
growing, community-maintained library of formal definitions, lemmas, theorems, and proof tools for
mathematics~\cite{mathlib}. 
Can we make one for data-system verification? 
Such a library might include:
\begin{itemize}[leftmargin=*,itemsep=2pt]
  \item %\textbf{Verified  protocols} at both the storage-engine and
% system levels. At the storage-engine level, these include optimistic lock
% coupling, latch crabbing, reader--writer locks, epoch-based reclamation, and
% lock-free retry loops. At the system level, they include protocols whose
% correctness yields transactional and replication guarantees, such as
% serializability, linearizability, snapshot isolation, and eventual consistency.
\textbf{Concurrency and coordination proofs} that capture recurring
  arguments about interactions among threads or nodes, such as consistent
  reads through optimistic validation, safe memory reclamation, and
  transaction ordering, with explicit assumptions for their reuse.
% hxc: This part too specific to verus mechanism, at least should belong to the 
%      related work if we want to keep them?
% These protocols can be specified as \emph{tokenized state machines} \cite{lattuada2024verus}, where a
% token is a proof-time abstraction of a thread's or process's permission to perform a logical
% step, such as accessing or modifying shared state. Tokens track the protocol's
% ownership and progress discipline: conceptually, they correspond to a lock in
% lock-based protocols and to a compare-and-swap in lock-free or
% optimistic protocols. Token exchanges encode transitions such as acquiring,
% transferring, consuming, or releasing a permission.

  \item \textbf{Representation and refinement lemmas} that connect concrete data structures to
  abstract models, % for example, 
  such as showing that a B$^+$ tree's nodes and pointers
  denote a well-formed ordered map.
  \item \textbf{Proof tactics and templates} that %guide the LLM 
  % from a failed verification obligation to the relevant 
  {help the agent identify suitable}
  invariants, %abstraction relation, 
  {decompose verification obligations,
  and instantiate} reusable lemma.
\end{itemize}

The central open question is how to design such a proof library so that it is
\emph{reusable without being rigid}. If its lemmas are too
implementation-specific, they will not generalize across designs; if they are
too abstract, they will do little to reduce the token cost of constructing a
proof. Can we identify a stable middle layer of database proof
knowledge---concurrency idioms, abstract-state refinements, and consistency
arguments---that enables an LLM to prove the correctness of a novel,
high-performance synthesized system without repeatedly rediscovering the same
semantic bridge?

\section{Looking Forward}
\label{sec:conclusion}

% Verifiable auto-research remains in its infancy. In this paper, we take a first look at how auto-research could be conducted for updatable data systems
% with consistency guarantees, treating token cost and research time as
% first-class citizens.  
% We are not alone in this pursuit. \cite{lao2026stop} shares our vision of verifiable auto-research and demonstrates its promise for read-only OLAP
% workloads. We bring this vision to updatable data
% systems with consistency guarantees, where concurrent correctness, distributed
% consistency, token cost, and research time become first-order research
% challenges.
% We believe that verifiable auto-research for data systems is a nascent area
% with many open problems for the community to tackle together.

Verifiable auto-research remains in its infancy. In this paper, we take a first
look at how auto-research could be conducted for data systems that are optimized
for general workloads, treating token cost and research time as 
first-class citizens. 

We are not alone in this pursuit.
IDS \cite{agarwal2026ids} shares our vision with respect to verifiability. Similarly, GenDB-V \cite{lao2026stop} and JetStream \cite{balasingam2026jetstream} respectively embrace this vision for OLAP workloads and only the read-side of a database, and do not consider research-time or token-consumption.
We bring this vision to updatable data systems, where correctness, token cost, and research time are first-order challenges. This setting creates numerous opportunities for community-wide investigation.

%% file: references_related_work.bib
@misc{balasingam2026jetstream,
  title         = {JetStream: Generating Verified Query Accelerators for Existing Database Systems},
  author        = {Balasingam, Akhilesh and Ngom, Amadou and Yu, Geoffrey X. and Kraska, Tim},
  year          = {2026},
  month         = {10},
  eprint        = {2610.20527},
  archivePrefix = {arXiv},
  url           = {https://arxiv.org/abs/2610.20527},
}

@article{amiri2026correctness,
  title   = {Correctness Firewalls for Synthesized Database Components},
  author  = {Amiri, Mohammad Javad and Binnig, Carsten},
  journal = {arXiv preprint arXiv:2611.05173},
  year    = {2026},
  url     = {https://arxiv.org/abs/2611.05173},
  note    = {Accessed November 9, 2026}
}

@inproceedings{ball2005abstraction,
  author    = {Ball, Thomas and Kupferman, Orna and Yorsh, Greta},
  title     = {Abstraction for Falsification},
  booktitle = {Computer Aided Verification: 17th International Conference, CAV 2005, Edinburgh, Scotland, UK, July 6--10, 2005, Proceedings},
  series    = {Lecture Notes in Computer Science},
  volume    = {3576},
  pages     = {67--81},
  year      = {2005},
  publisher = {Springer},
  doi       = {10.1007/11513988_8}
}

@inproceedings{lu2016snow,
  author    = {Lu, Haonan and Hodsdon, Christopher and Ngo, Khiem and Mu, Shuai and Lloyd, Wyatt},
  title     = {The {SNOW} Theorem and Latency-Optimal Read-Only Transactions},
  booktitle = {12th USENIX Symposium on Operating Systems Design and Implementation (OSDI 16)},
  year      = {2016},
  pages     = {135--150},
  publisher = {USENIX Association},
  url       = {https://www.usenix.org/conference/osdi16/technical-sessions/presentation/lu}
}

@inproceedings{konwar2021snow,
  author    = {Konwar, Kishori M. and Lloyd, Wyatt and Lu, Haonan and Lynch, Nancy},
  title     = {{SNOW} Revisited: Understanding When Ideal {READ} Transactions Are Possible},
  booktitle = {2021 IEEE International Parallel and Distributed Processing Symposium (IPDPS)},
  year      = {2021},
  publisher = {IEEE},
  doi       = {10.1109/IPDPS49936.2021.00101}
}

@article{jung2018iris,
  author  = {Jung, Ralf and Krebbers, Robbert and Jourdan, Jacques-Henri and Bizjak, Ale{\v{s}} and Birkedal, Lars and Dreyer, Derek},
  title   = {{Iris} from the Ground Up: A Modular Foundation for Higher-Order Concurrent Separation Logic},
  journal = {Journal of Functional Programming},
  volume  = {28},
  pages   = {e20},
  year    = {2018},
  doi     = {10.1017/S0956796818000151}
}

@misc{verusage,
  title         = {{VeruSAGE}: A Study of Agent-Based Verification for {Rust} Systems},
  author        = {Yang, Chenyuan and Neamtu, Natalie and Hawblitzel, Chris and Lorch, Jacob R. and Lu, Shan},
  year          = {2025},
  eprint        = {2512.18436},
  archivePrefix = {arXiv},
  primaryClass  = {cs.OS},
  url           = {https://arxiv.org/abs/2512.18436}
}

@misc{zhao2026perfreasoningllmsreasonhardware,
      title={PerfReasoning: How Well Do LLMs Reason on Hardware Performance?}, 
      author={Dan Zhao and Karthikeyan Sankaralingam and Christos Kozyrakis and Qijing Huang},
      year={2026},
      eprint={2609.04476},
      archivePrefix={arXiv},
      primaryClass={cs.AI},
      url={https://arxiv.org/abs/2609.04476}, 
}

@inproceedings{panickssery2024self,
  author    = {Panickssery, Arjun and Bowman, Samuel R. and Feng, Shi},
  title     = {{LLM} Evaluators Recognize and Favor Their Own Generations},
  booktitle = {Advances in Neural Information Processing Systems},
  volume    = {37},
  year      = {2024},
  doi       = {10.52202/079017-2197}
}

@article{wataoka2024self,
  author  = {Wataoka, Koki and Takahashi, Tsubasa and Ri, Ryokan},
  title   = {Self-Preference Bias in {LLM}-as-a-Judge},
  journal = {arXiv preprint arXiv:2410.21819},
  year    = {2024}
}

@article{khullar2026selfattribution,
  author  = {Khullar, Dipika and Hopkins, Jack and Wang, Rowan and Roger, Fabien},
  title   = {Self-Attribution Bias: When {AI} Monitors Go Easy on Themselves},
  journal = {arXiv preprint arXiv:2603.04582},
  year    = {2026}
}

@misc{lao2026stop,
  title         = {Stop Handcrafting Your Engine! GenDB-V: Towards AI-Generated and Formally Verified Database Systems},
  author        = {Lao, Jiale and Trummer, Immanuel},
  year          = {2026},
  month         = {11},
  eprint        = {2611.02779},
  archivePrefix = {arXiv},
}

@inproceedings{mathlib,
  author    = {{The mathlib Community}},
  title     = {The {L}ean {M}athematical {L}ibrary},
  booktitle = {Proceedings of the 9th {ACM} {SIGPLAN} International Conference on Certified Programs and Proofs},
  series    = {CPP 2020},
  publisher = {ACM},
  address   = {New Orleans, LA, USA},
  year      = {2020},
  month     = jan,
  doi       = {10.1145/3372885.3373824},
  url       = {https://doi.org/10.1145/3372885.3373824}
}

@article{abadi1991existence,
  author  = {Mart{\'i}n Abadi and Leslie Lamport},
  title   = {The Existence of Refinement Mappings},
  journal = {Theoretical Computer Science},
  volume  = {82},
  number  = {2},
  pages   = {253--284},
  year    = {1991},
  doi     = {10.1016/0304-3975(91)90205-P}
}

@article{gu2016certikos,
  author  = {Ronghui Gu and Zhong Shao and Hao Chen and Xiongnan (Newman) Wu and Jieung Kim and Vilhelm Sj{\"o}berg and David Costanzo},
  title   = {{CertiKOS}: An Extensible Architecture for Building Certified Concurrent {OS} Kernels},
  journal = {Proceedings of the 12th USENIX Symposium on Operating Systems Design and Implementation (OSDI)},
  year    = {2016},
  pages   = {653--669}
}

@inproceedings{lattuada2023verus,
author = {Lattuada, Andrea and Hance, Travis and Bosamiya, Jay and Brun, Matthias and Cho, Chanhee and LeBlanc, Hayley and Srinivasan, Pranav and Achermann, Reto and Chajed, Tej and Hawblitzel, Chris and Howell, Jon and Lorch, Jacob R. and Padon, Oded and Parno, Bryan},
title = {Verus: A Practical Foundation for Systems Verification},
year = {2024},
isbn = {9798400712517},
publisher = {Association for Computing Machinery},
address = {New York, NY, USA},
url = {https://doi.org/10.1145/3694715.3695952},
doi = {10.1145/3694715.3695952},
booktitle = {Proceedings of the ACM SIGOPS 30th Symposium on Operating Systems Principles},
pages = {438–454},
numpages = {17},
location = {Austin, TX, USA},
series = {SOSP '24}
}

@inproceedings{zelikman2024stop,
  author    = {Zelikman, Eric and Lorch, Eliana and Mackey, Lester and Kalai, Adam Tauman},
  title     = {Self-Taught Optimizer ({STOP}): Recursively Self-Improving Code Generation},
  booktitle = {Conference on Language Modeling (COLM)},
  year      = {2024}
}

@inproceedings{zhang2026darwin,
  author    = {Zhang, Jenny and Hu, Shengran and Lu, Cong and Lange, Robert and Clune, Jeff},
  title     = {Darwin {G}{\"o}del Machine: Open-Ended Evolution of Self-Improving Agents},
  booktitle = {International Conference on Learning Representations (ICLR)},
  year      = {2026}
}

@article{ferrari2025formal,
  author  = {Ferrari, Alessio and Spoletini, Paola},
  title   = {Formal Requirements Engineering and Large Language Models: A Two-Way Roadmap},
  journal = {Information and Software Technology},
  volume  = {181},
  pages   = {107697},
  year    = {2025}
}

@inproceedings{verifai,
  author    = {Beg, Arshad and O'Donoghue, Diarmuid and Monahan, Rosemary},
  title     = {Leveraging {LLMs} for Formal Software Requirements: Challenges and Prospects},
  booktitle = {Proceedings of the 7th International Workshop on Artificial Intelligence and
               Formal Verification, Logic, Automata, and Synthesis (OVERLAY 2025)},
  series    = {CEUR Workshop Proceedings},
  volume    = {4142},
  address   = {Bologna, Italy},
  year      = {2025}
}

@article{nishimuragasparian2026towards,
  author  = {Nishimura-Gasparian, Kei and McCarthy, Robert and Lindner, David},
  title   = {Towards Understanding Specification Gaming in Reasoning Models},
  journal = {arXiv preprint arXiv:2605.02269},
  year    = {2026}
}

@inproceedings{elmore2011zephyr,
  author    = {Elmore, Aaron J. and Das, Sudipto and Agrawal, Divyakant and El Abbadi, Amr},
  title     = {Zephyr: Live Migration in Shared Nothing Databases for Elastic Cloud Platforms},
  booktitle = {Proceedings of the 2011 ACM SIGMOD International Conference on Management of Data},
  pages     = {301--312},
  year      = {2011}
}

@article{das2011albatross,
  author  = {Das, Sudipto and Nishimura, Shoji and Agrawal, Divyakant and El Abbadi, Amr},
  title   = {Albatross: Lightweight Elasticity in Shared Storage Databases for the Cloud
             using Live Data Migration},
  journal = {Proceedings of the VLDB Endowment},
  volume  = {4},
  number  = {8},
  year    = {2011},
  doi     = {10.14778/2002974.2002977}
}

@article{bondarenko2025demonstrating,
  author  = {Bondarenko, Alexander and Volk, Denis and Volkov, Dmitrii and Ladish, Jeffrey},
  title   = {Demonstrating Specification Gaming in Reasoning Models},
  journal = {arXiv preprint arXiv:2502.13295},
  year    = {2025}
}

@inproceedings{zhong2026impossiblebench,
  author    = {Zhong, Ziqian and Raghunathan, Aditi and Carlini, Nicholas},
  title     = {{ImpossibleBench}: Measuring {LLMs'} Propensity of Exploiting Test Cases},
  booktitle = {International Conference on Learning Representations (ICLR)},
  year      = {2026}
}

@misc{penthalon,
      title={Decoupling Disaggregated Memory Optimizations from Indexing: A Compiler-Runtime Approach}, 
      author={Xinpeng Zhao and Zeling Long and Chaichon Wongkham and Srijan Srivastava and Jiayi Liu and Baotong Lu and Tianzheng Wang and Eric Lo},
      year={2026},
      eprint={2609.02669},
      archivePrefix={arXiv},
      primaryClass={cs.DB},
      url={https://arxiv.org/abs/2609.02669}, 
}

@misc{neuripsworkshop,
  author       = {Chen, Wuyang and Kong, Soonho and Oh, Hakjoo and He, Jingxuan and
                  Mitchell, Jacqueline and Liu, Amanda and Ye, Zhe and Liu, Xiaodong and
                  Pant, Varun and Lu, Jialin and Frieder, Simon},
  title        = {{AI} for Verifiable Coding: Human-Aligned Collaborative Agents for
                  Autoformalization, Proofs, and Heuristics},
  howpublished = {Workshop at the Conference on Neural Information Processing Systems (NeurIPS)},
  year         = {2026},
  url          = {https://neurips.cc/virtual/2026/workshop/137562}
}

@misc{specdb,
      title={SpecDB: LLM-Generated Customized Databases via Feature-Oriented Decomposition}, 
      author={Yunkai Lou and Longbin Lai and Shunyang Li and Zhengping Qian and Ying Zhang},
      year={2026},
      eprint={2605.31097},
      archivePrefix={arXiv},
      primaryClass={cs.DB},
      url={https://arxiv.org/abs/2605.31097}, 
}

@article{gendb,
author = {Lao, Jiale and Trummer, Immanuel},
title = {Demonstrating GenDB: Instance-Optimized and Customized Query Processing Code Generation via LLM Agents},
year = {2026},
issue_date = {August 2026},
publisher = {VLDB Endowment},
volume = {19},
number = {12},
issn = {2150-8097},
url = {https://doi.org/10.14778/3827998.3828076},
doi = {10.14778/3827998.3828076},
journal = {Proc. VLDB Endow.},
month = sep,
pages = {4602–4605},
numpages = {4}
}

@misc{alphaevolve,
      title={AlphaEvolve: A coding agent for scientific and algorithmic discovery}, 
      author={Alexander Novikov and Ngân Vũ and Marvin Eisenberger and Emilien Dupont and Po-Sen Huang and Adam Zsolt Wagner and Sergey Shirobokov and Borislav Kozlovskii and Francisco J. R. Ruiz and Abbas Mehrabian and M. Pawan Kumar and Abigail See and Swarat Chaudhuri and George Holland and Alex Davies and Sebastian Nowozin and Pushmeet Kohli and Matej Balog},
      year={2025},
      eprint={2506.13131},
      archivePrefix={arXiv},
      primaryClass={cs.AI},
      url={https://arxiv.org/abs/2506.13131}, 
}

@inproceedings{xu2026asiarch,
  author    = {Xu, Weixian and Liu, Yixiu and Nan, Yang and Ye, Lyumanshan and
               Hu, Xiangkun and Qin, Zhen and Liu, Pengfei},
  title     = {Neural Architecture Discovery via Autonomous Evolution},
  booktitle = {Conference on Language Modeling (COLM)},
  year      = {2026},
  url       = {https://proceedings.colm.cc/paper_files/paper/2026/file/0f6cc80ad86e553d085842308e0fd2cb-Paper-Conference.pdf}
}

@misc{openai2026navierstokes,
  author       = {{OpenAI}},
  title        = {Finite Time Blowup for {Navier--Stokes}},
  year         = {2026},
  month        = sep,
  howpublished = {\url{https://cdn.openai.com/pdf/32d9f210-8b73-45e0-91bc-82a30aef8a9a/navier-stokes.pdf}},
  note         = {Preprint. Accessed 2026-10-27}
}

@article{wehrstein2026bespoke,
  title   = {Bespoke OLAP: Synthesizing Workload-Specific One-Size-Fits-One Database Engines},
  author  = {Wehrstein, Johannes and Eckmann, Timo and Jasny, Matthias and Binnig, Carsten},
  journal = {Proceedings of the VLDB Endowment},
  volume  = {19},
  number  = {11},
  pages   = {3759--3771},
  year    = {2026},
  url     = {https://arxiv.org/abs/2603.02001}
}

@misc{jitskit,
  author       = {Liu, Shu and Krentsel, Alexander and Agarwal, Shubham and Cemri, Mert and Mao, Ziming and Ponnapalli, Soujanya and Dimakis, Alexandros G. and Ratnasamy, Sylvia and Zaharia, Matei and Parameswaran, Aditya and Stoica, Ion},
  title        = {The Time is Here for Just-in-Time Systems: Challenges and Opportunities},
  year         = {2026},
  eprint       = {2605.24096},
  archivePrefix= {arXiv},
  primaryClass = {cs.DB},
  url          = {https://arxiv.org/abs/2605.24096}
}

@inproceedings{skydiscover,
  author    = {Liu, Shu and Cemri, Mert and Agarwal, Shubham and Krentsel, Alexander and Naren, Ashwin and Mang, Qiuyang and Li, Zhifei and Gupta, Akshat and Maheswaran, Monishwaran and Cheng, Audrey and Pan, Melissa and Boneh, Ethan and Ramchandran, Kannan and Sen, Koushik and Zaharia, Matei and Dimakis, Alexandros G. and Stoica, Ion},
  title     = {SkyDiscover: A Flexible, Adaptive Framework for AI-Driven Scientific and Algorithmic Discovery},
  booktitle = {Proceedings of the ACM Conference on AI and Agentic Systems},
  series    = {CAIS '26},
  pages     = {1223--1227},
  year      = {2026},
  publisher = {Association for Computing Machinery},
  doi       = {10.1145/3786335.3813221},
  url       = {https://doi.org/10.1145/3786335.3813221}
}

@inproceedings{agarwal2026ids,
  author    = {Shubham Agarwal and Alexander Krentsel and Shu Liu and
               Mert Cemri and Audrey Cheng and Rui Meng and Tomas Pfister and
               Chun-Liang Li and Sylvia Ratnasamy and Aditya Parameswaran and
               Matei Zaharia and Ion Stoica and Mohsen Lesani},
  title     = {Inductive Deductive Synthesis: Enabling AI to Generate
               Formally Verified Systems},
  booktitle = {NeurIPS},
  year      = {2026},
}
